\documentclass[
 aps, pra,
 amsmath,amssymb,
 11pt,
 final,
tightenlines,
 twoside,
 twocolumn,
 nofloats,
nofootinbib,
 superscriptaddress,
showkeys,
showkeywords,
 ]
{revtex4-2}

\usepackage[T2A]{fontenc}
\usepackage[utf8x]{inputenc}
\usepackage[russian,english]{babel}
\usepackage{graphicx}
\usepackage{dcolumn}
\usepackage{bm}
\usepackage{soul}
\usepackage{xcolor}
\usepackage{geometry}
\input{maik.rty}
\setcitestyle{authoryear,round,aysep={,},citesep={;}}
\PassOptionsToPackage{authoryear}{natbib}
\def\squareforqed{\hbox{\rlap{$\sqcap$}$\sqcup$}}

\def\sq{\ifmmode\squareforqed\else{\unskip\nobreak\hfil
\penalty50\hskip1em\null\nobreak\hfil\squareforqed
\parfillskip=0pt\finalhyphendemerits=0\endgraf}\fi}

\def\utw{\smash{\rlap{\lower5pt\hbox{$\sim$}}}}

\def\udtw{\smash{\rlap{\lower6pt\hbox{$\approx$}}}}

\def\diameter{{\ifmmode\mathchoice
{\ooalign{\hfil\hbox{$\displaystyle/$}\hfil\crcr
{\hbox{$\displaystyle\mathchar"20D$}}}}
{\ooalign{\hfil\hbox{$\textstyle/$}\hfil\crcr
{\hbox{$\textstyle\mathchar"20D$}}}}
{\ooalign{\hfil\hbox{$\scriptstyle/$}\hfil\crcr
{\hbox{$\scriptstyle\mathchar"20D$}}}}
{\ooalign{\hfil\hbox{$\scriptscriptstyle/$}\hfil\crcr
{\hbox{$\scriptscriptstyle\mathchar"20D$}}}}
\else{\ooalign{\hfil/\hfil\crcr\mathhexbox20D}}%
\fi}}

\begin{document}

\selectlanguage{english}

\input engnames



\title{\Large{Photometric Stability of an EMCCD Camera at 1-s Exposures}} 

\author{\bf{\firstname{I.~V.}~\surname{Afanasieva}}}
\email{E-mail: \texttt{riv615@gmail.com}}
\affiliation{Special Astrophysical Observatory of the  Russian Academy of Sciences,
Nizhnii Arkhyz, 369167 Russia}

\author{\bf{\firstname{V.~G.}~\surname{Orlov}}}
\affiliation{Instituto de Astronom\'{\i}a, Universidad Nacional Aut\'onoma de M\'exico, 04510  M\'exico}

\author{\bf{\firstname{V.~I.}~\surname{Ardilanov}}}
\affiliation{Special Astrophysical Observatory of the  Russian Academy of Sciences,
Nizhnii Arkhyz, 369167 Russia} 

\author{\bf{\firstname{V.~A.}~\surname{Murzin}}}
\affiliation{Special Astrophysical Observatory of the  Russian Academy of Sciences,
Nizhnii Arkhyz, 369167 Russia}

\author{\bf{\firstname{D.~V.}~\surname{Oparin}}}
\affiliation{Special Astrophysical Observatory of the  Russian Academy of Sciences,
Nizhnii Arkhyz, 369167 Russia}

\author{\bf{\firstname{A.~N.}~\surname{Burenkov}}}
\affiliation{Special Astrophysical Observatory of the  Russian Academy of Sciences,
Nizhnii Arkhyz, 369167 Russia}
\received{August 19, 2024}  \revised{December 31, 2024} \accepted{January 25, 2025}

\begin{abstract}
\noindent {\bf{Abstact}}---We present the results of testing an \mbox{iXon Ultra 888} EMCCD camera to determine the operating parameters for short-exposure photometry of stars. As a result of the testing, those camera modes were selected in which the temporal instability of the electron multiplication charge does not significantly affect the light curves. In addition, the photometry of the eclipsing variable star ZTFJ\,0038+2030, obtained with the \mbox{Zeiss-1000} telescope of the Special Astrophysical Observatory of the  Russian Academy of Sciences, is presented. We have shown the advantages and disadvantages of 1-s exposures for studying variable stars.

\addvspace{0.3\baselineskip}

\noindent {\bf{DOI:}} \texttt{10.1134/S1990341324600832}

\addvspace{0.3\baselineskip}


\end{abstract}

\maketitle

\section{INTRODUCTION}

\vspace*{-0.8em}

The use of CCDs has \mbox{significantly expanded the} 
possibilities of observational astronomy (Howell,
2006). However, CCD-based cameras have one 
 \mbox{significant drawback:} they do not allow \mbox{fast readout of}
images with an acceptable noise level. \mbox{The cameras}
based on EMCCDs (Electron Multiplying Charge-Coupled 
Devices) do not have this disadvantage. 
\mbox{EMCCDs have} all the advantages of CCDs and also 
provide effective detection of single photons in low-light 
conditions by multiplying the charge in \mbox{a special} 
register (Robbins, 2011). At very low light fluxes 
\mbox{(less than} one photon per pixel per frame) \mbox{and using}
electron multiplication, the sensitivity of EMCCD 
cameras is significantly higher than that of the CCD
designs without charge multiplication (Rousset et al., 
2014). However, at a light flux \mbox{$F_l>1$\;photon/pixel} 
EMCCDs may exhibit unstable behavior.

At present EMCCD cameras are widely used in 
astronomy. They are used to obtain images with 
high angular resolution via ``lucky imaging'' (see, 
e.g., Schlagenhauf et al., 2024) and \mbox{speckle interferometry} 
(Maksimov et al., 2009; Mitrofanova et al., 2020; 
Orlov, 2021), to search for variable stars in \mbox{the central}
crowded regions of globular clusters (Skottfelt et al., 2015), to study the 
rapid variability of ZZ\,Cet-type stars in astroseismology 
(Angeles and Orlov, 2021), etc. A project 
\mbox{is underway} to use EMCCDs for the search and 
\mbox{spectroscopy of} exoplanets in the visible spectrum 
using the coronagraph on WFIRST (Wide Field 
\mbox{InfraRed Survey} Telescope).

Andor \mbox{iXon Ultra 888}\footnote{\url{https://andor.oxinst.com/products/ixon-emccd-cameras-for-life-science}}
is a versatile, highly 
sensitive, low-noise, and high-speed EMCCD camera.
\mbox{Under certain} conditions, its quantum efficiency can 
reach 95\%\ (Tulloch and Dhillon, 2011). However, it 
is very difficult to implement such a high \mbox{efficiency in} 
practice. The same is true for the electron multiplication 
(EM) gain $G_{\rm EM}$. In addition, many astronomers 
note unofficially the temporal instability of $G_{\rm EM}$. It
was noted that in some experiments the camera could 
not maintain the set temperature (Harps{\o}e et al., 
2012), which led to a drop in the EM gain, in some 
cases to one. 

The mentioned problems are the result \mbox{of incorrect} 
selection of camera parameters. \mbox{Andor Technology,} 
\mbox{the camera} manufacturer, provides the user with the 
ability to select parameters that are optimal for a 
\mbox{particular astronomical} task. \mbox{The manufacturer of} 
\mbox{the iXon Ultra 888} camera specifies the following 
\mbox{characteristics: a sensor} size of \mbox{$1024\times1024$} pixels, 
\mbox{a pixel size} of \mbox{$13\times13$\;$\mu$m}, and a maximum readout 
rate of 30\,MHz (26~frames per second). Only these 
camera characteristics are fixed. All \mbox{other camera} 
\mbox{features stated} by the manufacturer can be implemented 
with parameters specifically selected for an 
experiment. Detailed information about the detector 
characteristics of this camera can be found in Tulloch
and Dhillon (2011), Harding et al. (2015).

The sky background in the frame can be easily 
\mbox{subtracted using} classical procedures. However, the 
electronic bias and the dark current of the CCD are 
unstable. This paper examines only the hardware part, 
which is directly related to the camera design.

The main disadvantage of using the EM gain 
mode is the increase in image noise by a factor of 
\mbox{$\sqrt{2}$ (Robbins} and Hadwen, 2003). This parameter is
fundamental and does not depend on a selected $G_{\rm EM}$
value. Therefore, if the object is quite bright and does 
not allow setting high gain ($G_{\rm EM}>50$), it is more 
profitable to disable the EM gain.

If settings are selected correctly and the environment 
conditions do not vary, the camera would work 
fairly stably. However, during the observations one
may need to set other settings due to a change in
\mbox{environment parameters} or object characteristics. In
this paper, we analyze the time variation in camera 
bias and gain depending on selected parameters. It is
impossible to consider all possible scenarios of camera 
usage, so here we have limited ourselves to the 
\mbox{case of 1-s exposures}. We 
also describe the observing methodology and
present an analysis of the temporal stability of \mbox{iXon Ultra 888}
in photometric studies.

\vspace*{-0.5em}

\section{ADVANTAGES AND DRAWBACKS 
\mbox{OF PHOTOMETRY} WITH 1-S EXPOSURES}

\vspace*{-0.8em}

Due to the high EM gain the dynamic range of 
\mbox{EMCCDs is} tens or even hundreds of times smaller 
than that of conventional CCD detectors. Therefore, 
shorter exposures are necessary to avoid saturation.
The use of 1-s exposures has its own specific characteristics.

\vspace*{-1.2em}
\subsection{Disadvantages of Observing \mbox{with Short Exposures}}
\vspace*{-0.8em}

The main disadvantage of photometry with short
exposures is the need to store large amounts of data. 
However, with the decrease in the cost of external
drives this problem has lost its relevance.

Another disadvantage is that a short-exposure image
has a low signal-to-noise ratio (SNR), and the 
\mbox{main source} of the noise is photon (Poisson) noise
(Janesick, 2001). As a result, modern methods for
\mbox{constructing light} curves, such as difference image
analysis (DIA) (Bramich et al., 2015; Oelkers and
\mbox{Stassun, 2018),} require significant refinement.

The benefits of using short exposures easily outweigh 
these disadvantages.

\vspace*{-1.2em}
\subsection{Advantages}
\vspace*{-0.8em}

The evident advantages of observing with short 
\mbox{exposures include} a significant decrease of observation 
time waste caused by various factors. The most 
crucial of them are discussed below.

\vspace*{-0.9em}

\begin{list}{}{
\setlength\leftmargin{1mm} \setlength\topsep{3mm}
\setlength\parsep{0mm} \setlength\itemsep{8mm} }

\item[] {1. \it Cosmic Ray Hits.} \newline
The analysis of several of short-exposure images allows
better cleaning of the data from the traces of cosmic 
ray hits (CRHs). In most cases, after the cleaning
\mbox{procedure, no CRH} traces are left in the data, 
\mbox{since all trace} pixel values are below the quantization 
threshold.

\hspace*{1em}
In the case of very bright CRHs, automatic image
cleaning can be considered satisfactory \mbox{if the particle} 
trace does not affect the objects under study or comparison 
stars.

\hspace*{1em}
Unlike the standard CRH removal method described 
by Pych (2004), we construct a time histogram 
for each pixel. First, we look for \mbox{those pixels} 
\mbox{in which the histogram} (with an isolated peak) has
\mbox{a large gap} between the peaks. After that, in the 
frames where the pixel value is after the skip in the 
histogram, we replace it with the value before the gap.

\vspace*{-0.7em}

\item[] {2. \it Space Debris.}\newline
Recently, the amount of space debris in near-Earth 
space has increased so much that during an observing 
night several overflights can be recorded. Cleaning 
such images is very difficult and, as a rule, does 
\mbox{not give} satisfactory results.

\hspace*{1em}
In the case of observations with short exposures, 
only a few seconds of observing time are lost, which
\mbox{is an advantage} of using EMCCDs.

\vspace*{-0.7em}

\item[] {3. \it Telescope Guiding and Tracking Errors.}\newline
Due to errors of telescope tracking mechanisms 
or due to failures in automatic guiding 
during observations, the position of objects in the image 
shifts relative to the previous frame. In this case, 
all the objects in the summed or averaged image of a 
series of frames will look elongated or blurred.

\hspace*{1em} 
For observations with short exposures, image 
\mbox{offsets can be} easily compensated. 
To compensate guiding or tracking errors, we use the
centering technique described by Orlov et al. (2014).
Figure~1 shows, as an example, the averaged images of the 
Kepler-32f field before and after the correction of
tracking errors, panels (a) and (b), respectively.
\end{list}

\vspace*{-3.0em}
\section{STABILITY OF THE IXON ULTRA 888 CAMERA AT 1-S EXPOSURES}
\vspace*{-0.8em}

The main factor of camera stability is maintaining 
a constant temperature. We set the temperature 
\mbox{at $-60^\circ$C in} all laboratory experiments and $-80^\circ$C 
during telescope observations. The increase in temperature 
reduces the camera gain and increases the 
readout noise (Hirsch et al., 2013), therefore, the 
quality of data obtained from telescope observations
\mbox{is higher than} in the laboratory.

\begin{figure}
\includegraphics[scale=0.5]{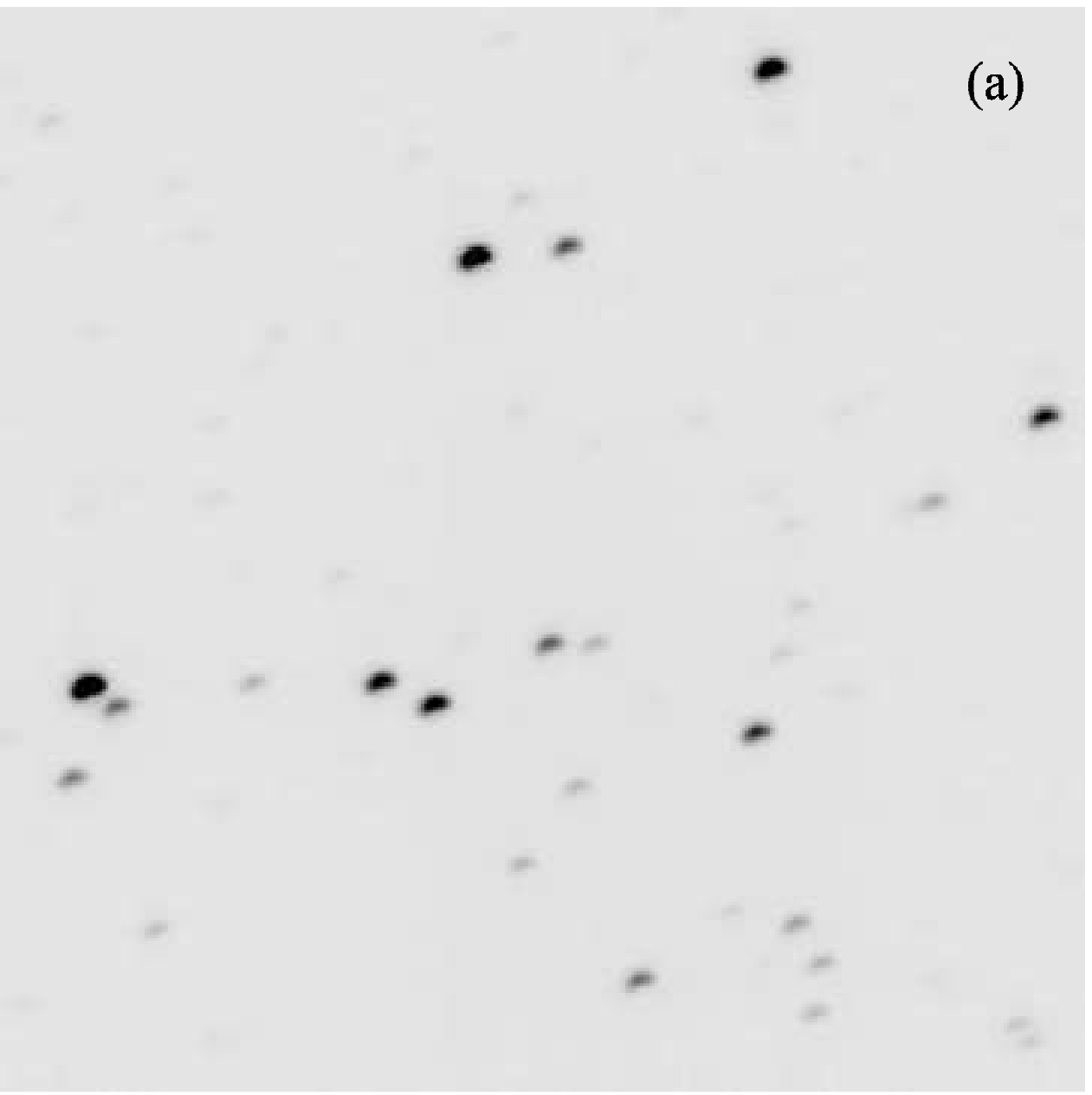}
\includegraphics[scale=0.5]{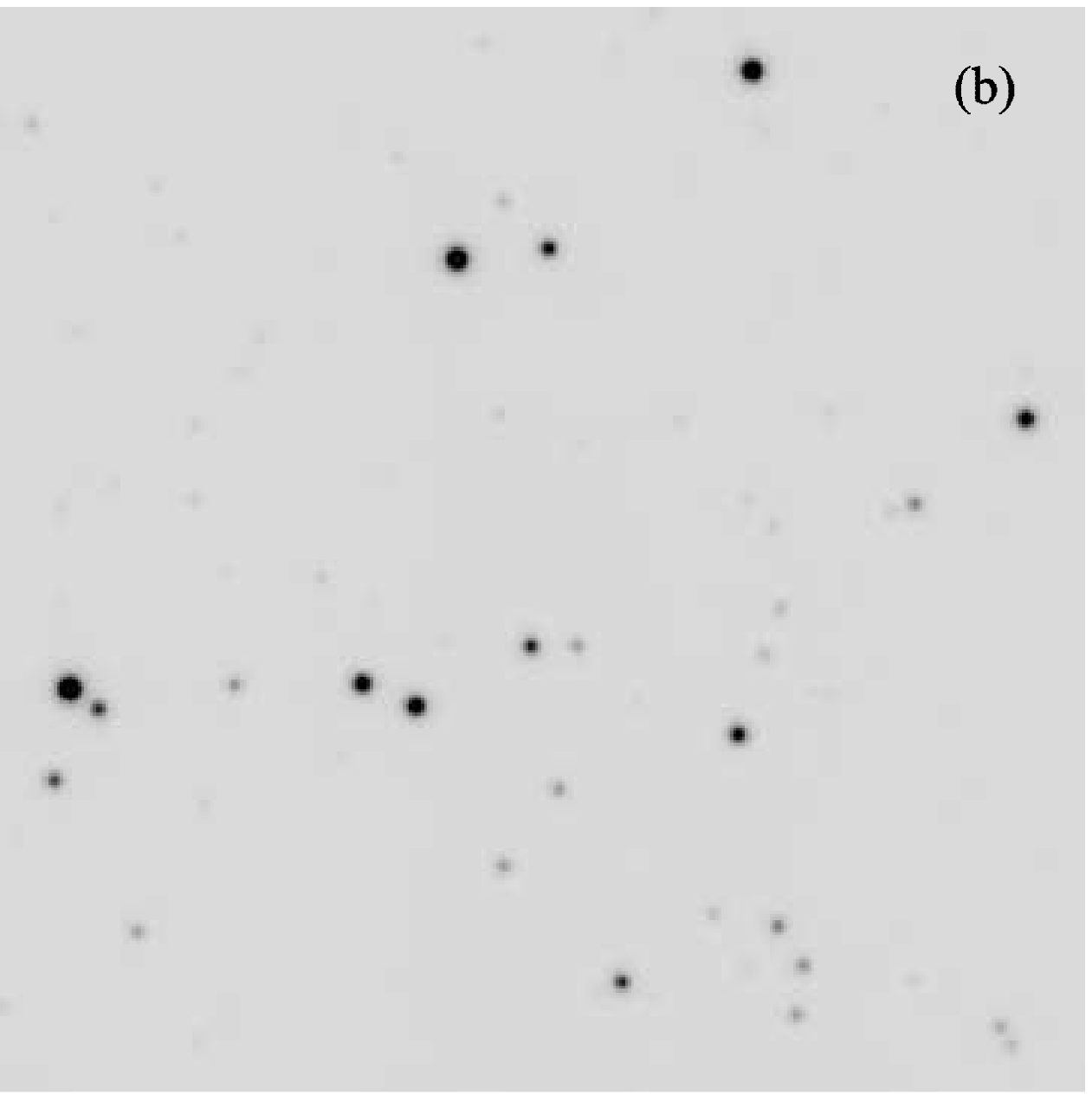}
\caption{Averaged images of a series of 1000~frames in 
\mbox{observations of Kepler-32f:} without correction of telescope
 tracking errors and after it---panels~(a) and (b), respectively. 
The displacement of objects in panel~(a) is $5\,.\!\!^{\prime\prime}5$ (9.6~pixels).}
\label{fig1}
\end{figure}

We studied the camera properties in the absence
of light and under uniform illumination in the spectral 
range of $450-700$\;nm with 1-s exposures using 
\mbox{flat-fielding}. To reduce the light flux to acceptable 
values, we placed a ThorLabs NE30A neutral density 
filter directly in front of the camera window using an 
adapter.

The data is collected in Non-Frame Transfer mode 
with different values of vertical shift speed (VSS) and 
horizontal readout rate (HRR) for two variants of 
\mbox{electron multiplication} of the signal before readout: 
with the gains $G_{\rm EM}=300$ and $G_{\rm EM}=1000$.

The VSS characterizes the time, in microseconds, 
required to shift the frame vertically by one line towards 
the shift register. The efficiency of charge 
transfer decreases at higher vertical shift speeds and 
leads to a deterioration in spatial resolution. The 
HRR characterizes the readout rate from the shift 
register.

Four vertical shift speeds (0.6, 1.13, 2.2, and 4.33\;$\mu$s) and
four readout frequencies (1, 10, 20,
\mbox{and 30\;MHz)} are available for this camera.~At 
\mbox{a speed of} 0.6\;$\mu$s, the frames demonstrate charge 
\mbox{leakage and} smearing even at a small light flux (about 
4000\;$\rm ADU$), therefore we excluded this mode from 
the study. The  1-MHz frequency was excluded
from consideration, because it does not meet the 
requirements of fast photometry.

\vspace*{-1.3em}
\subsection{Dark Image Analysis}
\vspace*{-0.8em}

To assess stability, we obtained a series of 1000 
\mbox{one-second} dark images with the shutter closed. 
These data allowed us to evaluate the behavior of 
\mbox{the camera at} low light fluxes (\mbox{$F_l<1$}~photon/pixel),
the stability of the shift over time, and the real gain 
(Daigle et al., 2006).

Tables 1 and 2 show the results of testing the 
\mbox{camera in} various modes with a gain of 300 and 1000, 
respectively. The last two columns of the tables show
temporal variation of the gain at different light fluxes.

The electronic offset depends on the gain in the 
\mbox{multiplication register} and the detector temperature. 
The offset consists of two parts: the fixed part, which 
is a geometric pattern that does not change during 
\mbox{observations, and} the variable part that does not depend 
on pixel coordinates but varies over time. The
variable part of the offset can be estimated as the truncated
mean of dark images. As shown in Fig.~2, the 
\mbox{offset has} small temporal variations with a standard 
deviation of about 0.58\;ADU. The camera readout 
noise can be estimated after removing the fixed part 
of the offset (Orlov, 2021).

\renewcommand{\baselinestretch}{0.8}
\begin{table}[!bpt] 
\caption{Temporal variation of the gain \mbox{$G_{\rm EM}=300$}} \label{tab1G}
\medskip
\begin{tabular}{c|c|c|c|c|c}
\hline
HRR,      & VSS,      & Real gain,    & \multirow{2}{*}{Err}     &  \multicolumn{2}{c}{Var, \%}  \\ \cline{5-6}
MHz &   $\mu$s &    $e^{-}$\,$\rm ADU^{-1}$ &        &  $F_l < 1$    &  $F_l = 9$   \\ \hline
(1) &   (2) &   (3) &(4) &  (5)&    (6)\\
\hline
     &  1.13     &  33.4     &  0.1  &  0.11     &  0.21    \\
10   &  2.2  &  32.6     &  0.1  &  0.10     &  0.19    \\
     &  4.33     &  32.8     &  0.1  &  0.10     &  0.18    \\
\hline
     &  1.13     &  26.3     &  0.2  &  0.22     &  0.43    \\
20   &  2.2  &  25.2     &  0.2  &  0.25     &  0.44    \\
     &  4.33     &  25.6     &  0.2  &  0.20     &  0.40    \\
\hline
     &  1.13     &  31.7     &  0.2  &  0.27     &  0.55    \\
30   &  2.2  &  28.3     &  0.2  &  0.27     &  0.58    \\
     &  4.33     &  28.6     &  0.1  &  0.28     &  0.54    \\
\hline
\end{tabular}
\end{table}

Figure~3 shows a histogram of dark frame values 
at \mbox{$G_{\rm EM}=1000$}, \mbox{$\rm VSS=2.2$}\;$\mu$s, \mbox{$\rm HRR=20$}\;MHz. The 
gain factor for this mode is about \mbox{$72\,e^{-}\,\rm ADU^{-1}$}.

A threshold processing scheme is used to distinguish 
single-photon events from false alarms caused 
by noise events. The threshold is set considering the 
average gain of electron multiplication (Basden et al., 
2003) and is about 27\;ADU.

The probability of charge multiplication in 
\mbox{EMCCDs} depends not only on the voltage at the 
electrodes of the multiplication register but also on the 
temperature of the sensor (Plakhotnik et al., 2006). 
The stability of these two factors affects the efficiency 
of electron multiplication. Reducing the sensor 
temperature by $70^\circ$\,C results in a tenfold increase in 
charge. To achieve high multiplication,  the sensor 
must be cooled. This camera uses a four-stage
Peltier element, which reduces the detector temperature by
80\,\textdegree C below ambient temperature. The main
characteristics of the EMCCD camera used are shown in Table~3.

Despite its high sensitivity at very low light levels,
EMCCD technology has a number of disadvantages. 
The gain mechanism necessary to reduce the 
\mbox{effective readout} noise below \mbox{$1e^-$}  also introduces an 
additional source of noise known as multiplicative 
\mbox{noise (Hynecek} and Nishiwaki, 2003; Robbins and 
Hadwen, 2003). This increases the shot noise of the 
signal by 1.41~times, resulting in greater variability 
\mbox{of the signal} in low light from pixel to pixel and from 
frame to frame.

\renewcommand{\baselinestretch}{1}
\renewcommand{\baselinestretch}{0.8}
\begin{table}[!h] 
\caption{Temporal variation of the gain \mbox{$G_{\rm EM}=1000$}} \label{tab2G}
\medskip
\begin{tabular}{c|c|c|c|c|c}
\hline
HRR,      & VSS,      & Real gain,    & \multirow{2}{*}{Err}     &  \multicolumn{2}{c}{Var, \%}  \\ \cline{5-6}
MHz &   $\mu$s &    $e^{-}$\,$\rm ADU^{-1}$ &        &  $F_l < 1$    &  $F_l = 9$   \\ \hline
(1) &   (2) &   (3) &(4) &  (5)&    (6)\\
\hline
     &  1.13     &  99.8     &  0.3  &  0.16     &  0.13    \\
10   &  2.2  &  98.3     &  0.2  &  0.15     &  0.11    \\
     &  4.33     &  99.6     &  0.2  &  0.15     &  0.11    \\
\hline
     &  1.13     &  75.7     &  0.2  &  0.31     &  0.21    \\
20   &  2.2  &  71.8     &  0.1  &  0.31     &  0.23    \\
     &  4.33     &  72.3     &  0.1  &  0.26     &  0.22    \\
\hline
     &  1.13     &  92.1     &  0.6  &  0.28     &  0.24    \\
30   &  2.2  &  83.3     &  0.3  &  0.28     &  0.26    \\
     &  4.33     &  83.7     &  0.3  &  0.30     &  0.21    \\
\hline
\end{tabular}
\end{table} \renewcommand{\baselinestretch}{1}

The cumulative effect of multiplicative noise is 
\mbox{seen in a} reduced SNR in the final image, which is 
statistically equivalent to halving the quantum efficiency 
of the camera (Tulloch and Dhillon, 2011). A 
\mbox{large pixel} size (from 13 to 16\;$\mu$m) can provide a high 
dynamic range, but only at a slower readout speed. Thus,
a high dynamic range can be attained with a lower frame rate
(or by reducing the frame size using binning) and low gain.

EMCCD cameras also have another significant 
\mbox{disadvantage---gain} degradation, which manifests at 
a certain temperature and voltage (Evagora et al., 2012).

\vspace*{-1.3em}
\subsection{Flat Field Analysis}
\vspace*{-0.8em}

To test the stability of the detector output signal 
\mbox{in the} laboratory, a series of 2000 one-second images 
were collected with uniform illumination by a constant 
light flux at a readout frequency of 20\;MHz and 
\mbox{a vertical shift} speed of 2.2\;$\mu$s.

\renewcommand{\baselinestretch}{0.9}
\begin{table}[b]
\caption {Characteristics of iXon Ultra 888} \label{tab1} \medskip
\begin{tabular}{l|c}
\hline
\multicolumn{1}{c|}{Parameter} & Value \\
\hline
Detector       & E2V CCD201-20  \\
Format, pixels      & $1024\times1024$   \\
Pixel size, $\mu$m    & $13\times13$   \\ 
Diagonal, mm       & 18.8  \\
EM readout noise, $e^{-}$   & < 1 @ $-$75\textdegree C     \\ 
Dark current,  $e^{-}$\,pixel$^{-1}$\,s$^{-1}$ 
& 0.00025 @ $-$80\textdegree C  \\ 
Operating temperature, \textdegree C    & $-55...-95$    \\ 
\hline
\end{tabular}
\end{table}%
\renewcommand{\baselinestretch}{1}

The first series of frames was obtained in the 
\mbox{``cold'' mode}, i.e., a lot of time had passed from the 
previous series: 545\;s (Fig.~4a). The second series 
\mbox{was received} 32\;s after the completion of the previous
one (Fig.~4b). It can be seen that in each of the modes, the signal
intensity decreases in the first 100--200\;s, 
and then reaches a steady saturation level with an
amplitude of variation of about 2--3\%.

This fact can be explained by the rise \mbox{and then fall} 
and stabilization of the EM gain. Gain variation can
be caused by several reasons. Firstly, by the instability 
of the detector temperature (at the beginning of the 
intensive readout, we recorded a decrease and then 
\mbox{an increase} in the detector temperature, see Fig.~4c).
Secondly, the instability of the high phase voltage 
\mbox{at the output} register of the detector.

These circumstances must be considered when 
\mbox{making photometric} measurements. Therefore, before 
the main observation of an object, it is recommended 
to readout the detector for at least \mbox{three to five} 
\mbox{minutes to} achieve a stable gain level. Additionally, there
should be no prolonged intervals between series of frames.

\vspace*{-1.3em}
\section{OBSERVATION \mbox{OF A ZTFJ\,0038+2030 ECLIPSE}}
\vspace*{-0.8em}

The observations were carried out on Novem-\mbox{ber~12, 
2022} using the Zeiss-1000 telescope of the 
\mbox{Special Astrophysical} Observatory (Komarov et al.,
2020) with a primary mirror of 1\;m in diameter and 
a focal length of $F/13.3$. To obtain direct images, 
\mbox{the camera} was set in the Cassegrain focus, and no 
\mbox{filters were} used. The diameter of the unvignetted 
\mbox{field of view} is $45^\prime$, the typical seeing for the North 
Caucasus weather conditions is about $1\,.\!\!^{\prime\prime}5$. The 
device is equipped with everything necessary for high-precision 
photometry of identified transient events.
\mbox{The field of view} is $3\,.\!\!^{\prime}46 \times 3\,.\!\!^{\prime}46$, the scale is $0\,.\!\!^{\prime\prime}2028$ 
per pixel for a frame without binning.

The main scientific purpose of the observations 
\mbox{with the Zeiss-1000} telescope was to test and refine
\mbox{a new technique} for observing fast-variability objects 
using an EMCCD camera and obtain the maximum
possible information about eclipsing stars.

\vspace*{-1.3em}
\subsection{ZTFJ\,0038+2030}
\vspace*{-0.8em}

To calibrate the measurement technique and determine
the accuracy of photometry, we carried out 
a photometric study of the eclipsing variable star
ZTFJ\,0038+2030 (Fig.~5) with known parameters.
This star is a white dwarf of almost 18th stellar 
\mbox{magnitude with} an invisible substellar component 
\mbox{(van Roestel} et al., 2021). The sky background was 
19~photons per pixel.

\begin{figure}[t]
\includegraphics[width=0.47\textwidth]{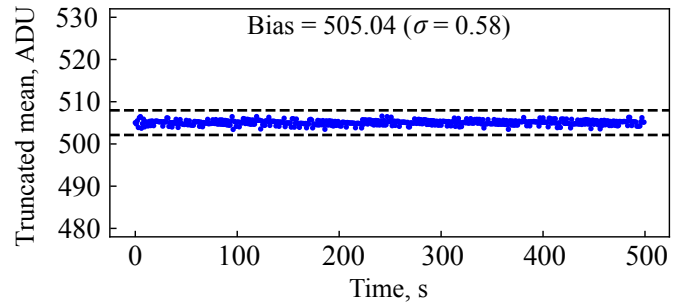}
\caption{A chart of bias in images. The average offset over 
the entire series is 505\;ADU, and the standard deviation 
\mbox{is about} 0.58\;ADU. The dashed lines delimit the $5\,\sigma$ 
\mbox{deviation from} the average.}
\label{fig2}
\end{figure}

\begin{figure}[t]
\includegraphics[scale=0.6]{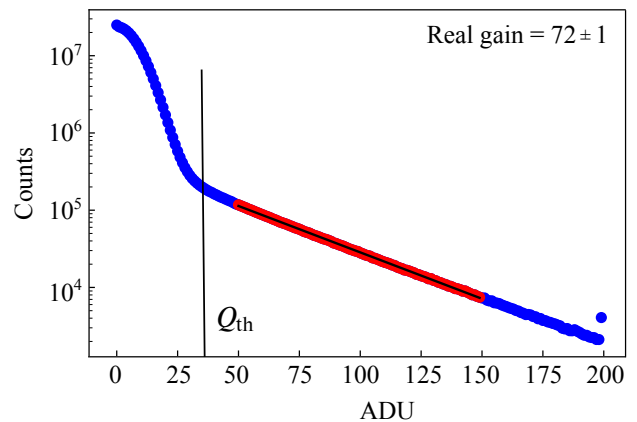}
\caption{The histogram of dark frame values is presented. 
The calculated values of the real gain and the quantization 
threshold are indicated.}
\label{fig3}
\end{figure}

The following criteria were applied to the choice of
the object: the presence of an eclipse during observations, 
the object altitude not lower than $60^\circ$ above 
\mbox{the horizon}, and the maximum angular distance from
the Moon. With a readout rate of 20\;MHz and an 
accumulation time of 1\;s, the exposure time of one
frame in a kinetic series (KCT, kinetic cycle time) 
was 1.06526\;s. Since the object is very faint, we 
chose the maximum gain $G_{\rm EM}=1000$. The real
 gain (conversion factor) for the selected parameters 
was \mbox{$K=71.8\;e^{-}\,\rm ADU^{-1}$}. In total, three series of 
1000~frames were received.

\vspace*{-1.3em}
\subsection{Data Processing}
\vspace*{-0.8em}

\begin{figure*}[t]
\includegraphics[scale=0.8]{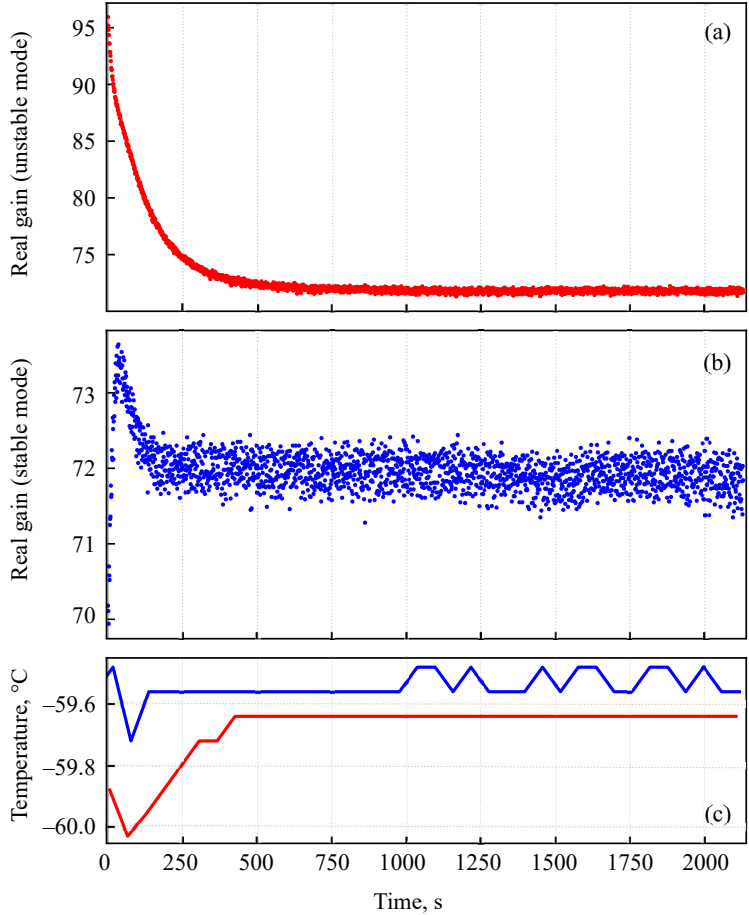}
\caption{Gain factor variation depending on time at $G_{\rm EM}=1000$,
$\rm VSS=2.2$\;$\mu$s, and $\rm HRR=20$\;MHz. The graph in panel~(a)
shows the typical behavior of the gain in unstable camera operation,
the graph on panel~(b) is for the gain variation in steady-state mode.
Panel~(c) shows the temperature variation during the gain measurements 
in the unstable (red line) and stable (blue line) camera modes.}
\label{fig4}
\end{figure*}

When the flux $F_l$ exceeds five photons per pixel per 
frame, an EMCCD behaves like a typical low-noise 
CCD. In such cases, the flux in a pixel is determined 
by subtracting the dark-frame value and dividing the 
result by the real gain at which the image was obtained
(Daigle et al., 2009):
\begin{equation*}
I(x,y,t) = (I'(x,y,t)-D(x,y))/K,
\end{equation*}
where $I'(x,y,t)$ is the intensity of an original image 
number $t$ in the data cube at a point with coordinates
($x, y$), $I(x,y,t)$ is the intensity of the corrected 
image, $D(x,y)$ is a robust evaluation of a series of 
dark images at a point with coordinates ($x, y$), and
$K$ is  the conversion factor (real gain of the camera).
For \mbox{$F_l<5$\;photon/pixel}, the optimal transformation 
\mbox{is that} of Rousset et al. (2014):
\begin{equation*}
I(x,y,t) =  \begin{cases}
{\rm int}[(I'(x,y,t)-D(x,y))/K], \\ \quad \text{when } I'(x,y,t) \geq Q_{\rm th}, \\
0, \text{when } I'(x,y,t) < Q_{\rm th},
        \end{cases}
\end{equation*}
where $\rm int[]$ is the integer part, $Q_{\rm th}$ is the threshold
obtained from a histogram of dark images (see Fig.~3).

The next step is flat field correction. We have to 
perform this procedure for each frame due to gain 
\mbox{variation. After} that, we corrected the telescope
\mbox{tracking errors}.

\begin{figure*}[t]
\includegraphics[scale=0.48]{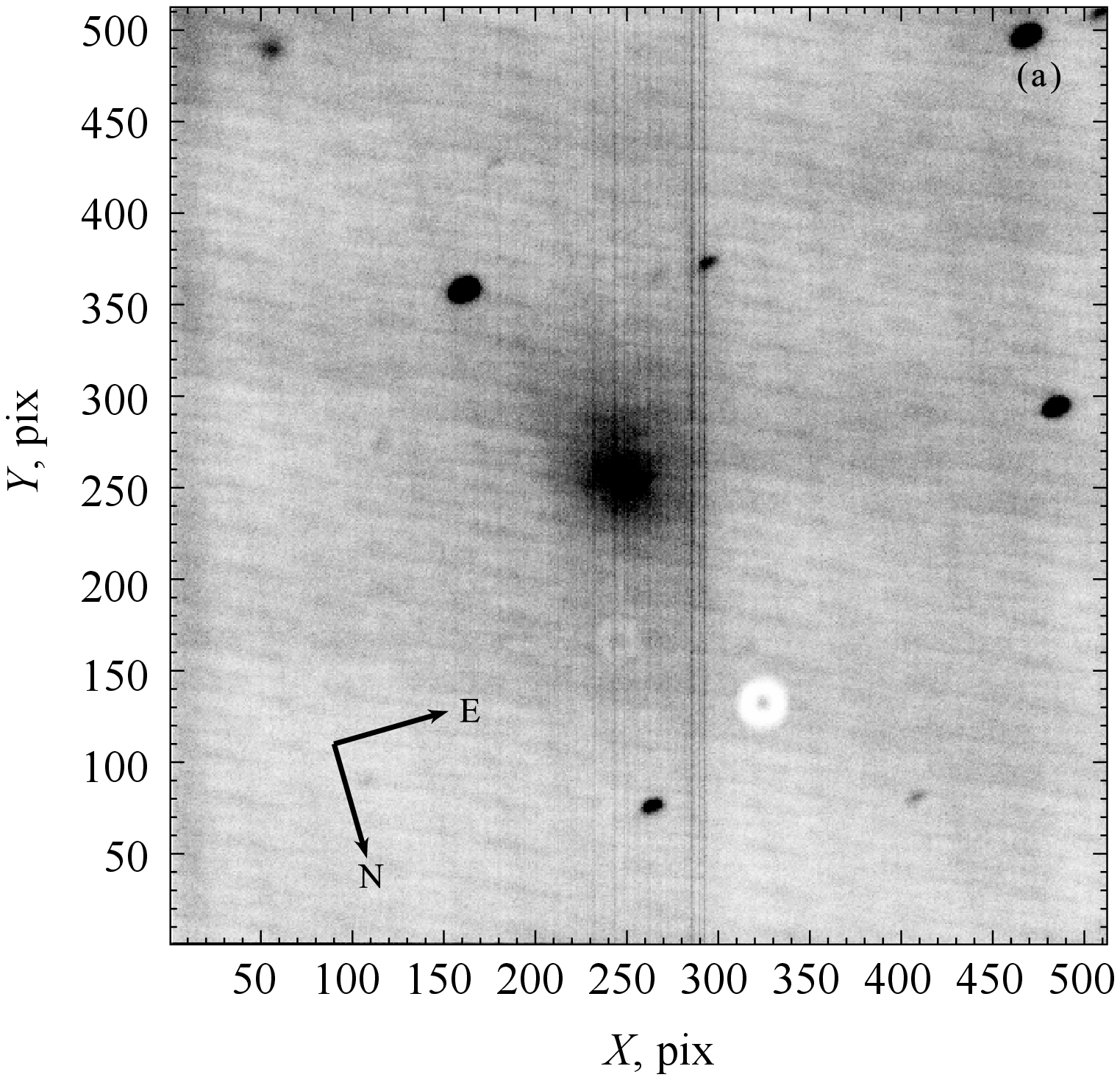} \hspace{-0.8cm} 
\includegraphics[scale=0.48]{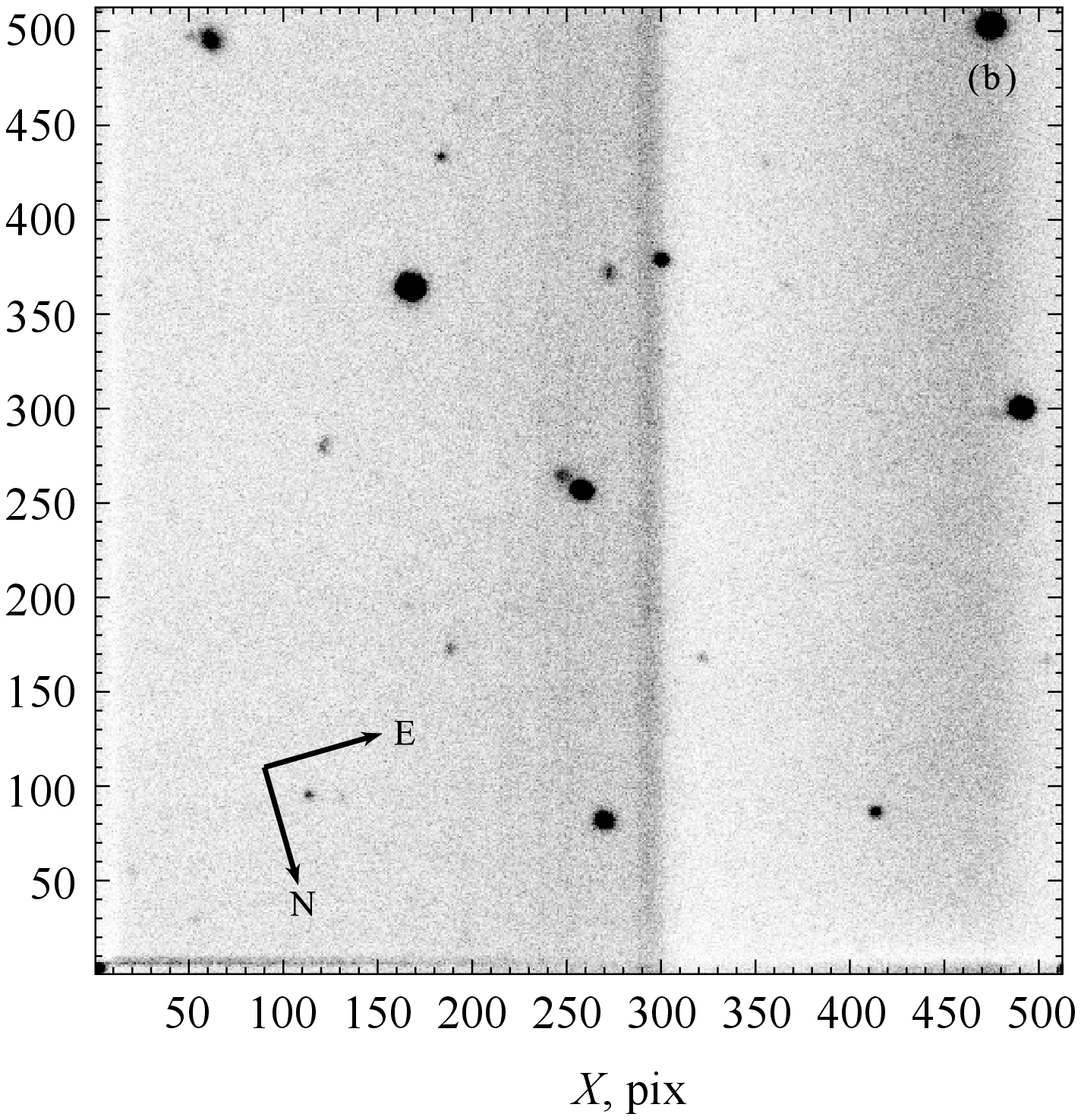} 
\caption{An image of the eclipsing variable star ZTFJ0\,0038+2030 (in the center), obtained by summing 1000 frames: before~(a)
and after (b) the processing: dark subtraction, flat field correction, removal of telescope guiding errors, and CHR cleaning.
The frame size is \mbox{$3\,.\!\!^{\prime}46 \times 3\,.\!\!^{\prime}46$}.}
\label{fig5}
\end{figure*}

The final preprocessing stage involves removing
\mbox{cosmic ray} hits from the images. Since after the 
correction for the errors of telescope guiding the signal 
in the image ceases to obey the \mbox{Poisson distribution} 
(Labb\'e et al., 2003), weak traces of cosmic rays 
\mbox{cannot be removed} by constructing time diagrams for
each pixel. In this regard, it is recommended to use 
alternative cleaning methods or perform the CRH removal
procedure before correcting the guiding errors.

After the initial processing, the \mbox{data can be}
\mbox{processed using} standard packages designed for 
\mbox{photometric data} analysis. We used the package
{\tt Photutils} (Bradley et al., 2024) affiliated with the
{\tt Astropy} library (Price-Whelan et al., 2022). An open
source {\tt Python} package {\tt Photutils} provides tools for
detection and photometry of astronomical sources.

The light curve of a ZTFJ\,0038+2030 eclipse with
1-s resolution is shown in Fig.~6. If there are no short-term
events in the light curve, the exposure time can be 
increased by adding frames.

\begin{figure*}[t]
\includegraphics[scale=0.90]{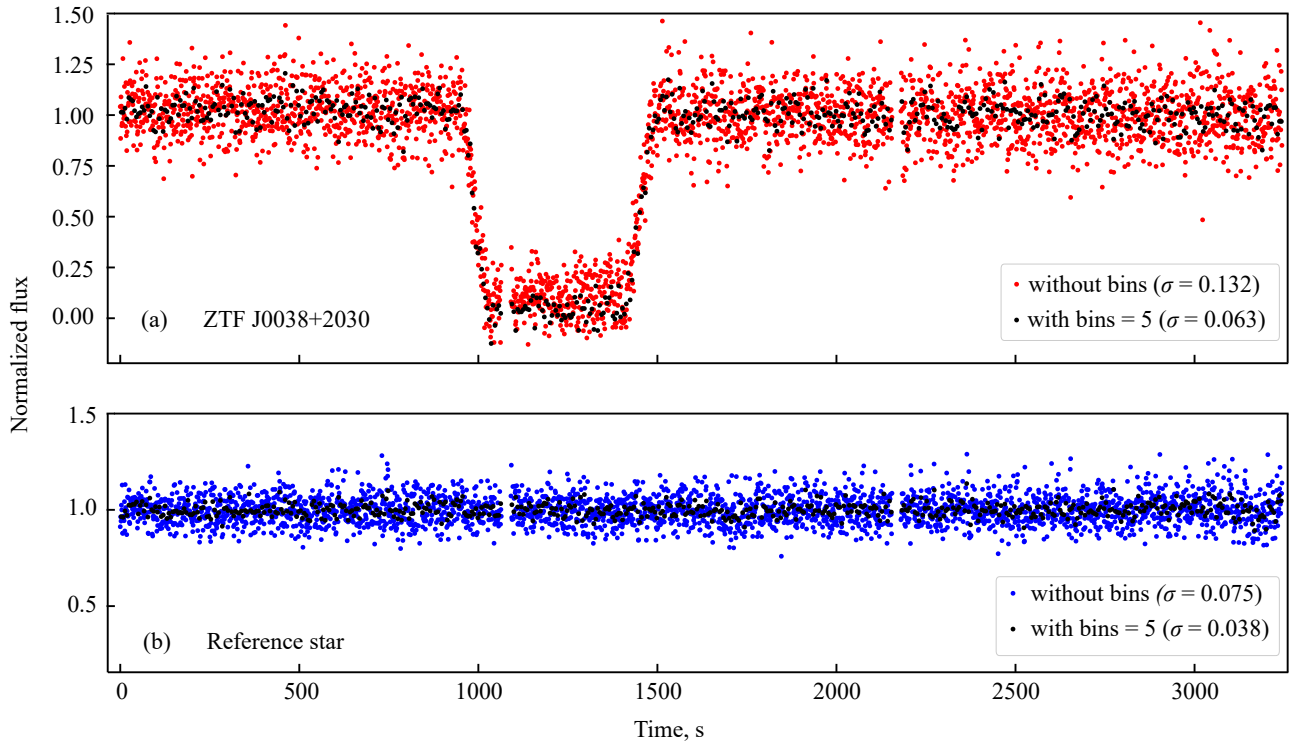}
\caption{The light curves of ZTFJ\,0.0038+2030 (a) and a comparison
star (b). The plots show the measurement results without a preliminary
summing of the data (colored dots) and with the preliminary summing of
five frames before processing (black dots).}
\label{fig6}
\end{figure*}

\vspace*{-1.2em}
\section{RESULTS AND DISCUSSION}
\vspace*{-0.8em}

The actual camera gain values we obtained were
\mbox{significantly lower} than those stated by the manufacturer. 
This is due to the fact that our experiment used
a temperature higher than required. \mbox{To obtain higher}
gain values, the temperature should \mbox{be reduced} to
the maximum permissible level. Implementing liquid
\mbox{cooling will} be necessary to maintain such a temperature, 
as it is the only way to expect a real gain close 
to the declared value.

In addition, it is necessary to set the camera pre-gain 
value to 2. However, even under these conditions, 
at high light fluxes it is unlikely that it is going
to be possible to maintain the real gain at a level close
to that stated by the manufacturer. For example, with 
the temperature set to $-70$\textdegree\;C and a pre-gain of 2,
the camera stabilized at a real gain of \mbox{110\;$e^{-}\,\rm ADU^{-1}$};
and at $-60$\textdegree\;C and a pre-gain of 1, it stabilized at only 
\mbox{30\;$e^{-}\,\rm ADU^{-1}$}.

We believe that strong temperature reduction
\mbox{without liquid} cooling is not always justified, as the 
\mbox{camera may not} be able to maintain such a temperature
during operation. This may cause the camera 
to malfunction or even turn off. We recommend using
liquid cooling as a guarantee of stable operation at all
ambient temperatures. Without it, the real gain of the
camera may be lower than stated.

The camera enables efficient astronomical 
observations with short exposures, even with low real
\mbox{gain. However}, in this case it is almost impossible
\mbox{to achieve} the accuracy required to study exoplanets
with high temporal resolution.

\vspace*{-1.2em}
\section{CONCLUSION}
\vspace*{-0.8em}

Photometry with 1-s exposures allow us \mbox{to obtain}
precise light curves by removing telescope guiding 
\mbox{errors and} by correcting frames corrupted by the 
overflight of satellites or by cosmic ray hits.


The EMCCD iXon Ultra 888 camera is well suited
for such observations since after reaching stable camera
mode it provides sufficient amplification of the light 
flux. The variation of the gain factor in stable
camera mode does not exceed 0.3\% and 0.5\% with a 
gain set to 1000 and 300, respectively. The time to
reach stable mode depends on the selected gain and
the intensity of the light flux. As a rule, it does not
exceed 10 min even with a light flux of more than
30~photons per pixel.

Observations with 1-s exposures of the eclipsing
variable star ZTFJ\,0038+2030 (magnitude of about
$18^{\rm m}$) using a 1-m telescope allowed us to obtain 
the light curve of its eclipse with a standard deviation 
of 13.2\%. Increasing the exposure by summing five 
frames reduced the light curve error to 6.3\%.

\vspace*{-1.0em}
\begin{acknowledgments}
\vspace*{-0.8em}
Observations with the telescopes \mbox{of the Special}
Astrophysical Observatory, Russian Academy of Sciences
\mbox{(SAO RAS),} are supported by the Ministry
of Science and Higher Education of the Russian
\mbox{Federation. This paper} used the catalogs of the SIMBAD 
database (https://simbad.u-strasbg.fr/simbad/) and 
Encyclopaedia of Exoplanetary Systems (https://
exoplanet.eu/).
\end{acknowledgments}

\vspace*{-1.0em}
\section*{FUNDING}
\vspace*{-0.8em}

The work was performed as part of the SAO RAS 
government contract approved by the Ministry of 
\mbox{Science and} Higher Education of the Russian Federation.
This research is also supported by the Direcci{\'o}n
General de Asuntos del Personal Acad{\'e}mico (UNAM,
M{\'e}xico) under project IN114123.

\vspace*{-1.0em}
 \section*{CONFLICT OF INTEREST}
\vspace*{-0.8em}
The authors of this work declare that they have no
conflict of interest.


\vspace*{5mm}

\hspace*{35mm}\it{Translated by D. Kudryavtsev}



%
\onecolumngrid

\end{document}